\documentclass[preprint,showpacs,preprintnumbers,amsmath,amssymb]{revtex4}
\usepackage{dcolumn}% Align table columns on decimal point
\usepackage{bm}
\usepackage{amssymb,amsmath, amsthm,epsfig}

\begin{document}

\title{Cascades in decaying three-dimensional electron magnetohydrodynamic turbulence}
\author{CHRISTOPHER J. WAREING\thanks{Email: cjw@maths.leeds.ac.uk} and RAINER HOLLERBACH}
\affiliation{Department of Applied Mathematics, University of Leeds, Woodhouse Lane, Leeds, LS2 9JT, UK}

\begin{center}
Article submitted to Journal of Plasma Physics 28th March 2009. 

Accepted 1st June 2009. Article will appear in revised form, subsequent to editorial
input by Cambridge University Press, in Journal of Plasma Physics
(http://journals.cambridge.org/PLA)
\end{center}

\begin{abstract}
Decaying electron magnetohydrodynamic (EMHD) turbulence in three dimensions 
is studied via high-resolution numerical simulations.  The resulting energy 
spectra asymptotically approach a $k^{-2}$ law with increasing $R_B$,
the ratio of the nonlinear to linear timescales in the governing equation,
consistent with theoretical predictions. 
No evidence is found of a dissipative cutoff, consistent with non-local 
spectral energy transfer and recent studies of 2D EMHD turbulence. 
Dissipative cutoffs found in previous studies are explained as 
artificial effects of hyperdiffusivity. 
In another similarity to 2D EMHD turbulence, relatively stationary 
structures are found to develop in time, rather than the variability 
found in ordinary or MHD turbulence.  Further, cascades of energy in
3D EMHD turbulence are found to be suppressed in all directions under 
the influence of a uniform background field.  Energy transfer is further
reduced in the direction parallel to the field, displaying scale
dependent anisotropy. 
Finally, the governing equation is found to yield a weak inverse cascade, 
at least partially transferring magnetic energy from small to large scales.
\end{abstract}

Copyright 2009 Cambridge University Press. This article may be downloaded
for personal use only. Any other use requires prior permission of the author
and of Cambridge University Press.\\

\maketitle

\section{Introduction}\label{sec:intro}

Turbulence plays a crucial role in a wide variety of geophysical and
astrophysical fluid flows.  In this paper we present results on a
specific variety of plasma turbulence in which the flow consists entirely 
of electrons, moving through a static background of ions.  The
equation governing the electrons' self-induced magnetic field is then
\begin{equation}
\frac{\partial {\bf B}}{\partial t} =
 - \nabla \times \left[ {\bf J} \times {\bf B} \right] 
+ R_B^{-1} \ \nabla^2 {\bf B},
\label{eq:A}
\end{equation}
where ${\bf J} = \nabla \times {\bf B}$, and $R_B={\sigma B_0}/{n e c}$,
with $\sigma$ the conductivity, $B_0$ a measure of the field strength,
$n$ the electron number density, $e$ the electron charge, and $c$ the
speed of light.  See for example \cite{goldreich92}, who derived this
equation in the context of magnetic fields in the crusts of neutron
stars.  More generally though, it is applicable in many weakly
collisional, strongly magnetic plasmas, so other applications could
include the Sun's corona or the Earth's magnetosphere.

Turbulence governed by (\ref{eq:A}) is known as electron MHD (EMHD), Hall MHD,
or whistler turbulence.  Based on its (at least superficial) similarity
to the vorticity equation governing ordinary, nonmagnetic turbulence,
\begin{equation}
\frac{\partial {\bf w}}{\partial t} =
\nabla \times \left[ {\bf u} \times {\bf w} \right]
 + Re^{-1} \ \nabla^2 {\bf w},
\label{eq:B}
\end{equation}
where now ${\bf w} = \nabla \times {\bf u}$, \cite{goldreich92}
argued that (\ref{eq:A}) would initiate a turbulent cascade to small
lengthscales, thereby accelerating neutron stars' magnetic field
decay beyond what ohmic decay acting on large lengthscales could
achieve.  They suggested in particular that the turbulent spectrum
would scale as $k^{-2}$, with a dissipative cutoff occurring at
$k\sim R_B$.

However, recent work has highlighted the fundamental differences 
between equations (\ref{eq:A}) and (\ref{eq:B}) (\cite[Wareing 
and Hollerbach 2009]{wareing09}, henceforth referred to as WH09).  
In (\ref{eq:B}) the dissipative term contains more derivatives 
than the nonlinear term, so on sufficiently short lengthscales 
the dissipative term will always dominate and hence there is a 
dissipative cutoff.  In contrast, in (\ref{eq:A}) the two terms 
both contain two derivatives, so it is conceivable that the 
nonlinear term will always dominate, even on arbitrarily short 
lengthscales. \cite[WH09]{wareing09} found precisely this effect in 
decaying 2D EMHD turbulence.

In this paper we present high-resolution numerical simulations of
(\ref{eq:A}) in a three-dimensional (3D) periodic box geometry, 
designed specifically to address such questions as to whether there 
is a dissipative cutoff or not, and whether the coupling is local 
or not.  In contrast to previous 3D simulations 
(\cite[Biskamp et al. 1999, Biskamp and M\"uller 1999, Cho and 
Lazarian 2004]{biskamp99,biskamp99b,cho04}), we do not employ hyperdiffusivity, 
which disrupts this feature that the two terms 
in (\ref{eq:A}) have the same number of derivatives, and hence 
introduces an artificial dissipative cutoff (\cite[WH09]{wareing09}).
These previous simulations found no difference between 2D and 3D
turbulent spectra, with both having a $k^{-7/3}$ scaling.
In our recent study of the governing equation in two dimensions
with normal diffusivity (\cite[WH09]{wareing09}), we found no evidence for a dissipative 
cutoff, and the turbulent spectrum scaled as $k^{-5/2}$, broadly 
consistent with previous results. We now
consider whether 3D EMHD turbulence displays the same scaling 
characteristics. Finally, we also consider the 
question of whether (\ref{eq:A}) is capable of yielding an inverse 
cascade in 3D (we use hyperdiffusivity for this set of runs).

\section{Numerics}\label{sec:equations}

We solve (\ref{eq:A}) by treating the $z$-independent parts of $\bf B$ as
before (\cite[WH09]{wareing09}).  For the $z$-dependent parts we expand $B_x$ and
$B_y$ in triple Fourier series in $x$, $y$ and $z$, with $B_z$ then given
by $\nabla\cdot{\bf B}=0$. Time integration is achieved through a second order
Runge-Kutta method, with the diffusive term treated exactly. We employ standard
pseudospectral techniques for the evaluation of the nonlinear terms, with
dealiasing according to the 2/3 rule. The code employs the MPI 
library\footnote{http://www.mcs.anl.gov/mpi/} and the 
FFTW library (\cite[Frigo and Johnson 2005]{frigo05}) 
to achieve massive parallelisation on a suitable supercomputer. 
We performed a variety of runs, typically employing 64
processors, with the highest extending to $k=170$ in Fourier space,
corresponding to $N=512$ collocation points in real configuration 
space. Due to the two derivatives in the nonlinear term, the 
required timesteps are unfortunately very small, roughly 
proportional to $1/(N^2)$.  Values as small as $\sim1\times 10^{-6}$ 
were used, requiring $O(2 \times 10^5)$ timesteps to reach $t=0.2$.

\subsection{Initial Conditions}

Since our interest is in freely decaying, rather than forced,
turbulence, we need to carefully consider the nature of our chosen
initial conditions.  We will present results for three different
sets of runs.

First, to study homogeneous forward cascades, we start off with
random $O(1)$ energies in all Fourier modes up to $k=
(k_x^2+k_y^2)^{1/2}=5$, making sure that the energy is evenly 
distributed between the three components.
After initialisation the overall amplitude 
of the field is rescaled to ensure that the rms value of 
$|{\bf B}|=1$ at $t=0$.

Second, to study nonhomogeneous forward cascades, we start off with
the same initialisation as above, but now add a uniform field
$C\hat{\bf e}_x$, where $C=1$, 2, 4 or 8.  This field is simply 
added to the $B_x$ component directly, resulting in a suitably 
modified equation (\ref{eq:A}).

Third, to explore the possibility of inverse cascades, we return to
the $C=0$ case without any large scale magnetic field and now inject 
energy into modes in the range $10\le k\le 20$.  The question then is 
how much of this initial energy moves to $k<10$, and how much moves 
to $k>20$.  

Finally, for all three sets of results, each individual run was
repeated with a number of different random initial conditions, to
ensure that the results presented here are indeed representative.

\subsection{Ideal Invariants}

Equation (\ref{eq:A}) has some useful associated diagnostics, 
corresponding to quantities that are conserved in the ideal, 
$R_B^{-1} -> 0$, limit.  Specifically, in three dimensions we 
have equations for the energy,
\begin{equation}
\frac{d}{dt} { \frac{1}{2} \int {\bf B}^2\,{\rm d}V } 
= -R_B^{-1} \int {\bf J}^2\,{\rm d}V,
\label{q1}
\end{equation}
and the magnetic helicity,
\begin{equation}
\frac{d}{dt}{\frac{1}{2}\int{\bf A\cdot B}\,{\rm d}V}
=-R_B^{-1}\int{\bf B\cdot J}\,{\rm d}V,
\label{q2}
\end{equation}
where $A$ is the vector potential, defined by ${\bf B=\nabla\times A}$.
Note though that in the presence of a uniform background field,
helicity is not even defined (\cite[Berger 1997]{berger97}), let alone conserved.

Whereas in 2 dimensions, we also had the additional quantity 
of the mean squared magnetic potential or anastrophy, this is
not conserved in 3 dimensions. This quantity is particularly important
for the inverse cascade in 2D EMHD, where such a cascade is thought to
be driven by a forward cascade of energy and an inverse cascade of
anastrophy (\cite[Shaikh and Zank 2005]{shaikh05}). It is unclear then, 
whether an inverse cascade will occur in 3D.

In addition to the physical insight that these various integrated
quantities yield into the nature of the Hall nonlinearity, they also
offer useful diagnostic checks of the code.  Reassuringly, we found
that both of them (except helicity in a uniform field) were
satisfied to within 0.25\% or better by all of our runs.

\section{Results}\label{sec:results}

\subsection{Large-scale initial conditions}

A characteristic statistical quantity of a turbulent system is the
energy spectrum. 3D hydrodynamic turbulence exhibits the spectrum
$E_k \propto k^{-5/3}$, the famous Kolmogorov law (\cite[Kolmogorov 1941]{kolmogorov41}).
In MHD turbulence, the energy transfer is altered by the Alfv\'en
effect (\cite[Iroshnikov 1964, Kraichnan 1964]{iroshnikov64,kraichnan65}), leading to a 
flatter energy spectrum $E_k \propto k^{-3/2}$.
Recent studies of 2D EMHD turbulence have found, via methods which
all employ hyperdiffusivity, a 5/3 Kolmogorov spectrum for small 
scales $k d_e > 1$, equivalent to $k > O(R_B)$, and a steeper 7/3 
spectrum for longer wavelengths 
(\cite[Biskamp et al. 1996, Biskamp et al. 1999, Dastgeer et al. 
2000, Dastgeer and Zank 2003, Cho and Lazarian 2004, Shaikh and 
Zank 2005]{biskamp96,biskamp99,dastgeer00,dastgeer03,cho04,shaikh05}).
Of the studies which considered 3D EMHD turbulence, \cite{biskamp99} 
numerically found a scaling of 7/3 for $k d_e < 1$ consistent with a
local spectral energy transfer independent of the linear wave
properties. However, the authors of that paper noted the scaling
in 3D was only marginally verified, since the short extent
of the inertial spectral range was dominated by the bottleneck
effect, causing local enhancement of the spectrum
above the inertial range power law at the point of transition from
the inertial range to the dissipation range. The more abrupt the
transition (or the higher the degree of hyperdiffusivity), the more
pronounced the effect.

\begin{figure}
\includegraphics[angle=0]{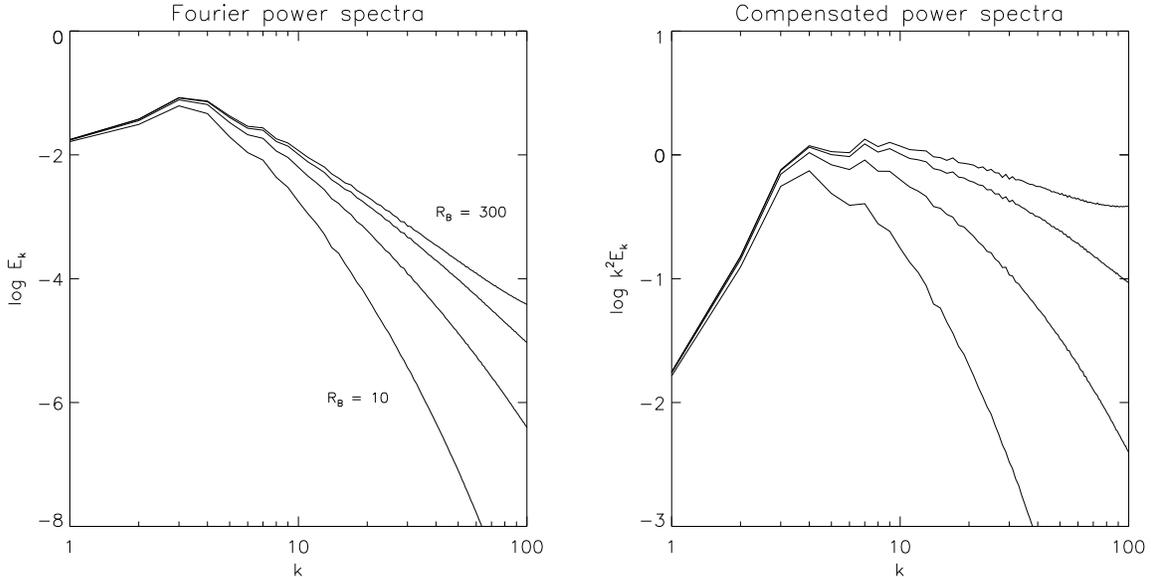}
\caption{Energy spectra of homogeneous 3D EMHD turbulence at 
$t=0.2$. On the left are shown spectra for $R_B=10$, 30, 100, 
\& 300. On the right, compensated spectra $k^{2} E_k$ 
for the same range of $R_B$.}
\label{cascades}
\end{figure}

In the left plot of Figure \ref{cascades}, we show the energy 
spectra of our solutions for $R_B = 10$, 30, 100 \& 300, evolved 
to a time $t = 0.2$. The energy spectra have been stationary since 
approximately $t = 0.16$ and time averaging between 0.16 and 0.2 
reveals an identical spectrum and no further information. We 
interpret this to mean our simulations are resolved and evolved 
to a suitable time for inspection of the now quasi-stationary 
cascade. It is worth noting at this point that 3D EMHD turbulence 
takes a longer time to reach quasi-stationarity than 2D EMHD
turbulence, which achieves such a phase by $t\approx 0.1$.

The energy spectra all start out much the same at low $k$, peak around $k=3$ and 
then lower $R_B$ spectra smoothly drop off with increasing $k$ 
whilst higher $R_B$ spectra maintain a linear gradient in the 
log-log plot. Transfer of energy to higher $k$ is then more 
efficient at higher $R_B$, with $R_B = 300$ having a scaling of $\sim k^{-5/2}$. 
The spectra are asymptotically approaching an energy spectrum 
$E_k \propto k^{-\nu}$, where we propose $\nu \sim 2$. In 
the right plot of Figure \ref{cascades}, we show compensated 
energy spectra to show this approach to $k^{2}E_k = 1$ with
increasing $R_B$. Our value of $\nu$ is the same as that
predicted by \cite{goldreich92} for 3D EMHD turbulence. Their 
prediction was calculated using a phenomenology based on Kraichnan's 
arguments (the whistler effect) and is reproduced in \cite{biskamp96}.
\cite{biskamp96} also considered a theoretical prediction neglecting
the whistler effect and derived a scaling of $7/3$. Simulations
have shown the whistler effect has little effect on the energy 
spectrum of 2D EMHD turbulence, but it would seem that it may 
have a considerable effect on 3D EMHD turbulence. 

Regarding the bottleneck effect, we see no evidence for it in the 
spectra. This is not a great surprise as we see no sign of a
dissipative cutoff, which causes the effect. By definition, 
the dissipation scale should occur when the local value of $R_B$ 
is $O(1)$ in equation (\ref{eq:A}). It is unclear though when this 
occurs since the definition of $R_B$ does not involve length 
scales. If the coupling is purely local in wavenumber, then this 
definition does involve length scales after all, since the $B_0$ 
that should be used is the field at that wavenumber only, rather 
than the total field. That is, according to the definition of 
\cite{hollerbach02} where this argument was first developed, we 
have
\begin{equation}
R_B' = R_B (B'/B)
\end{equation}
where the primed quantities are the small-scale local values and
the unprimed the large-scale global. If we now suppose a $k^{-2}$ 
energy spectrum, then $B'/B \sim k^{-1}$ and so $R_B'$ is 
reduced to $O(1)$ when $k \sim R_B$. So, at $R_B = 100$ we 
would see a dissipative cutoff at $k \sim 100$, which we do not.
It is possible to reconcile the situation by realising
that this argument crucially depends on the coupling being local 
in Fourier space: if this does not hold then $R_B' = R_B$ and 
there is simply no definite dissipation scale.

In agreement then with our 2D study of EMHD turbulence, the nonlinear 
term is able to dominate at all length scales and the coupling is 
therefore non-local in Fourier space. Again, hyperdiffusivity has 
previously masked the effect of the nonlinear term at high $k$. 
It has also introduced the bottleneck effect which has disguised
the true scaling of the energy spectra. With normal diffusivity,
we believe we have performed the first true simulations of 3D
EMHD turbulence.

\begin{figure}
\includegraphics[angle=0]{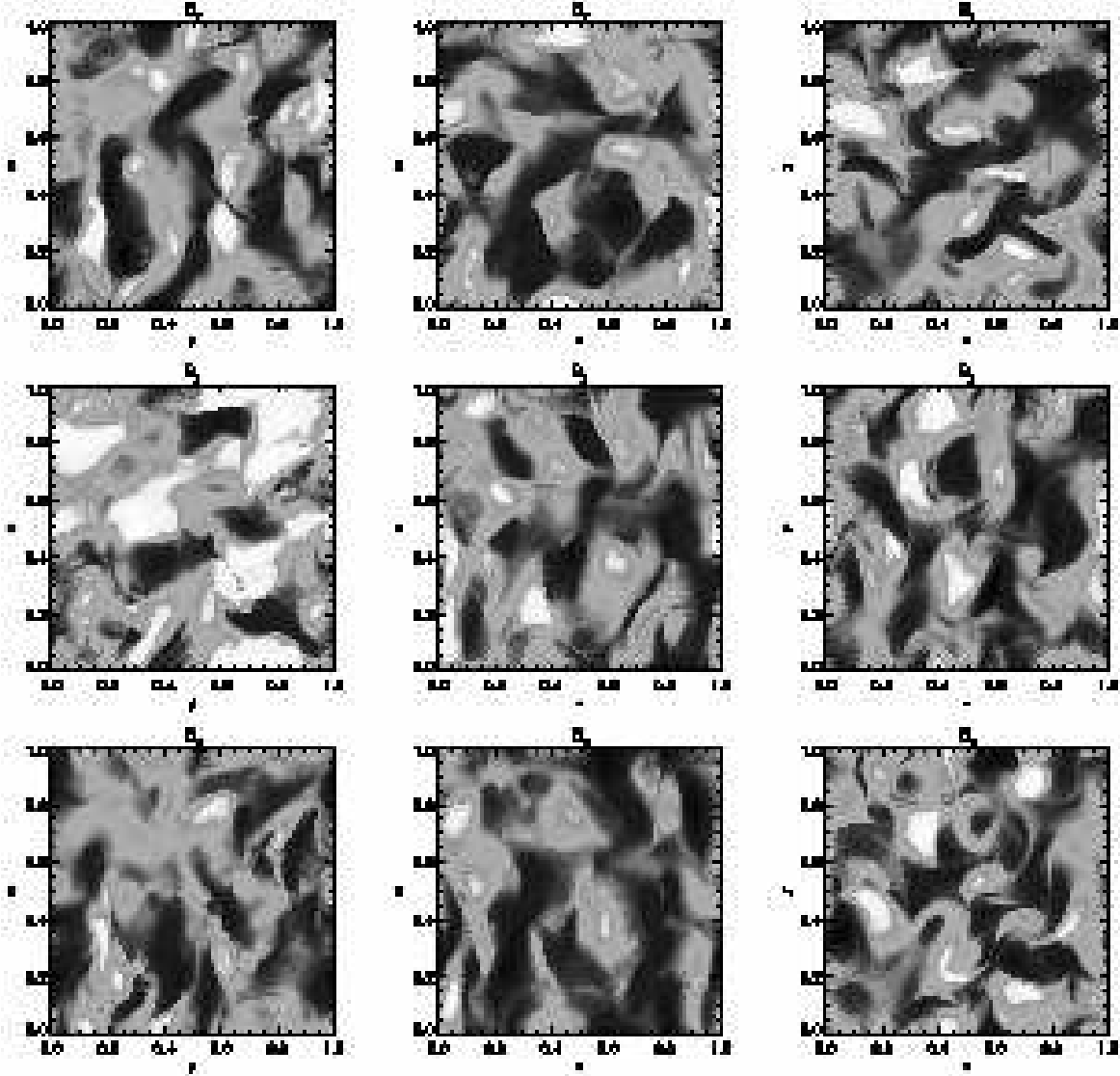}
\caption{Plots of the $R_B = 300$ solution in real
configuration space at $t = 0.1$. For full
details of the slices through the datacube, see the text.}
\label{truefield-0.1}
\end{figure}

\begin{figure}
\includegraphics[angle=0]{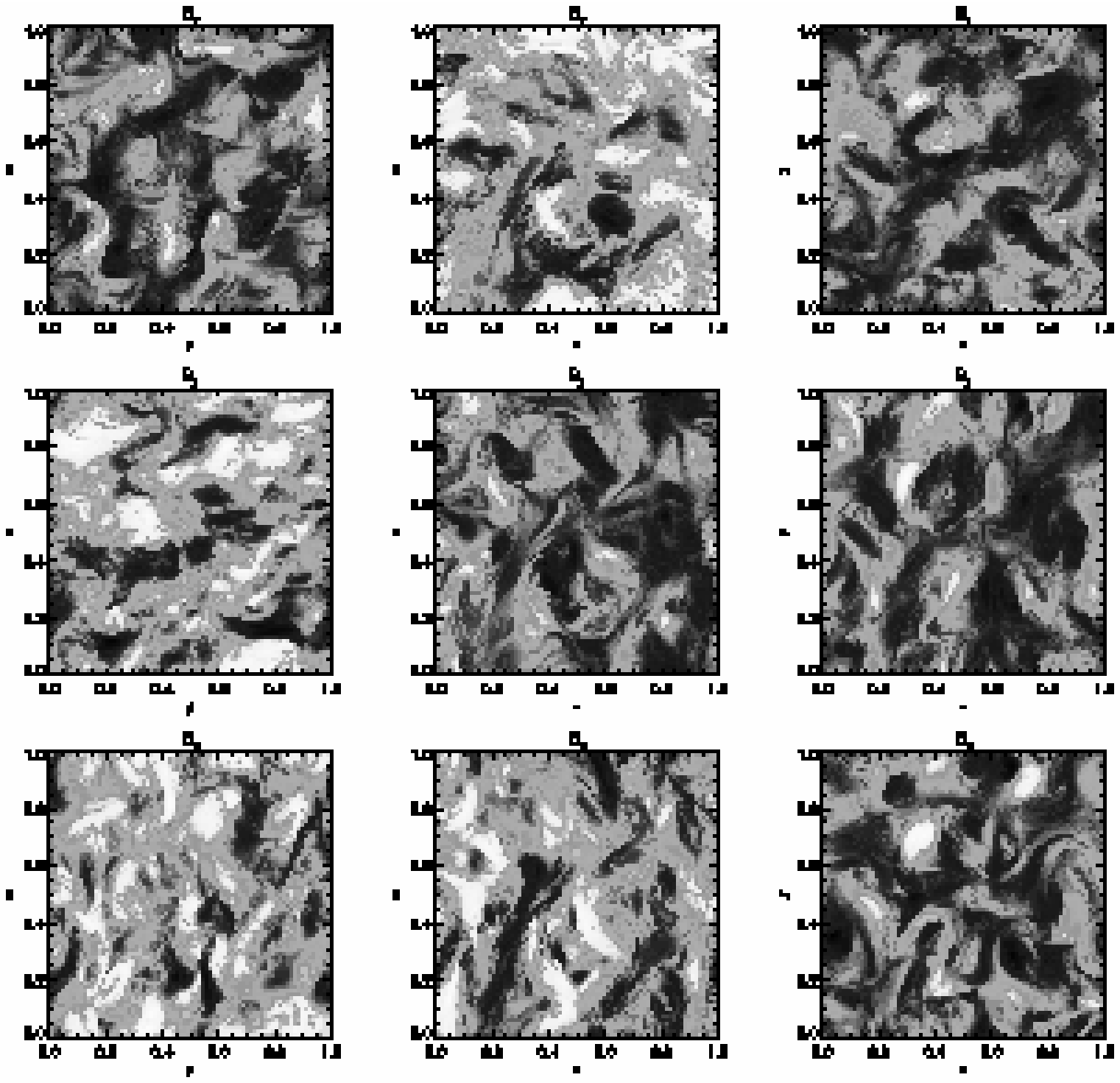}
\caption{Plots of the $R_B = 300$ solution in real
configuration space at $t = 0.2$. For full
details of the slices through the datacube, see the text.}
\label{truefield-0.2}
\end{figure}

In Fourier space then, EMHD turbulence continues to bear a strong
resemblance to ordinary MHD turbulence. We would like to know
if this resemblance carries over into real configuration space. In
Figure \ref{truefield-0.1}, we show slices through the three component 
field datacubes at $t = 0.1$. Across the top row, we show a slice
of the $B_x$ field at $x = 0.5$ on the left, a slice at $y = 0.5$ 
in the middle and a slice at $z = 0.5$ on the right. Across the 
middle row we show the same slices for $B_y$ and across the bottom
row for $B_z$. Large numbers of small, independent vortices dominate
the fields, characteristic of fully developed turbulence.

In Figure \ref{truefield-0.2}, we show the same fields at $t=0.2$. It 
is clear that the fields resemble those at $t=0.1$, implying 3D 
EMHD turbulence is much more structured than classical and MHD 
turbulence, where a fully developed turbulent field would bear no 
resemblance to the initial field. This appears to be a unique 
characteristic of decaying EMHD turbulence, in both two and three 
dimensions.

\begin{figure}
\includegraphics[angle=0]{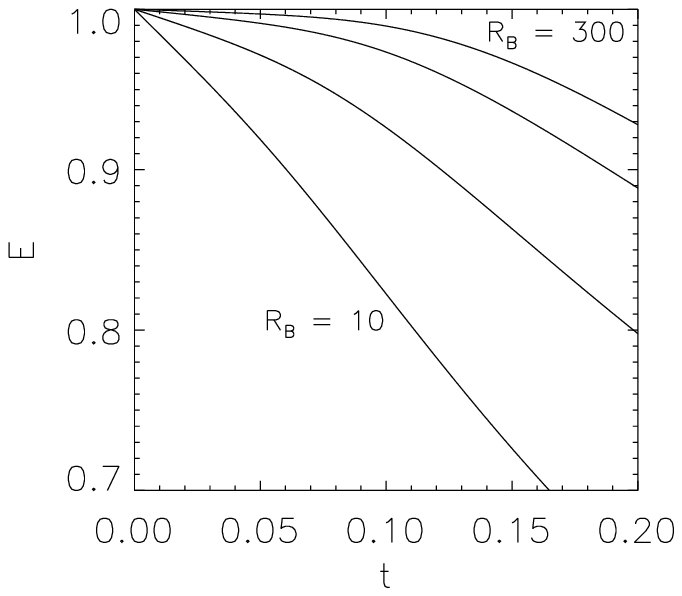}
\caption{A plot of energy against time for $R_B=10$, 30, 100 \& 300,
increasing from left to right.}
\label{energy}
\end{figure}

We would also like to address the energy decay of the field, with
particular respect to any dependency of the decay rate on the
value of $R_B$. \cite{biskamp99b}  
reported that the energy dissipation rate is independent of the
value of the dissipation coefficient, represented by $R_B$ here.
In contrast, \cite[WH09]{wareing09} found the energy decay is strongly
dependent on the $R_B$ parameter, with the decay rate proportional
to $R_B^{-1}$. We find the same behaviour here, as shown 
in Figure \ref{energy}.

\subsection{Large-scale initial conditions in the presence 
of a background field}

3D EMHD turbulence, like 2D EMHD, classical and MHD 
turbulence, is isotropic when allowed to freely decay. In the 
presence of a background flow, classical turbulence remains 
isotropic. Small-scale structures are advected along by any 
large-scale flow, whether or not that has a uniform 
background contribution. This effect has been attributed to
local coupling in phase space. Numerical simulations of MHD 
turbulence have found it to be strongly anisotropic in the 
presence of a background field (\cite[Shebalin et al. 1983, 
Oughton et al. 1998]{shebalin83,oughton98}). 
This has been attributed to the excitation of Alfv{\' e}n 
waves which preferentially propagate parallel to the external 
magnetic field and hinder the cascade process perpendicular 
to the external field.

In 2D EMHD turbulence, recent numerical studies employing
hyperdiffusivity (\cite[Dastgeer et al. 2000, Dastgeer and 
Zank 2003]{dastgeer00,dastgeer03}) have revealed 
anisotropic behaviour with the cascade strongly inhibited 
parallel to the background field. This can only be the result of
asymmetry in the nonlinear spectral transfer process relative
to the external magnetic field. In the context of local energy 
coupling in Fourier space, mediation by whistler waves has
been proposed as the only way this asymmetry could be achieved
(\cite[Dastgeer et al. 2000]{dastgeer00}), by the mechanism 
detailed by \cite{galtier06}. 
Most recently, the numerical study of \cite[WH09]{wareing09} has
confirmed the anisotropy of 2D EMHD with normal diffusivity.
The spectrum of 2D anisotropic EMHD turbulence has also been 
shown to exhibit a linear relationship with an external 
magnetic field (\cite[Dastgeer and Zank 2003]{dastgeer03}). 

\begin{figure}
\includegraphics[angle=0]{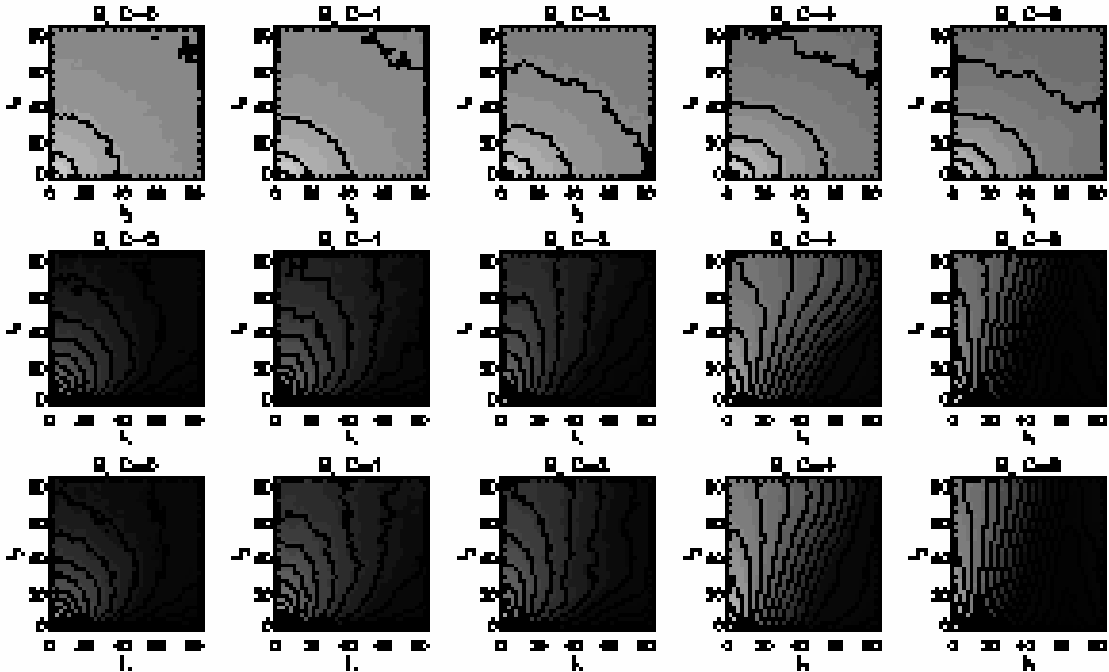}
\caption{2D Fourier power spectra at $t=0.2$ for 3D EMHD turbulence 
in the presence of a background field. We show three 2D Fourier power 
spectra slices of the $B_x$ datacube, at $k_x = 1$ (top row), $k_y = 1$
(middle row) and $k_z = 1$ (bottom row). Across the columns we show
the power spectra for $C = 0$, 1, 2, 4 \& 8.}
\label{background1}
\end{figure}

\begin{figure}
\includegraphics[angle=0]{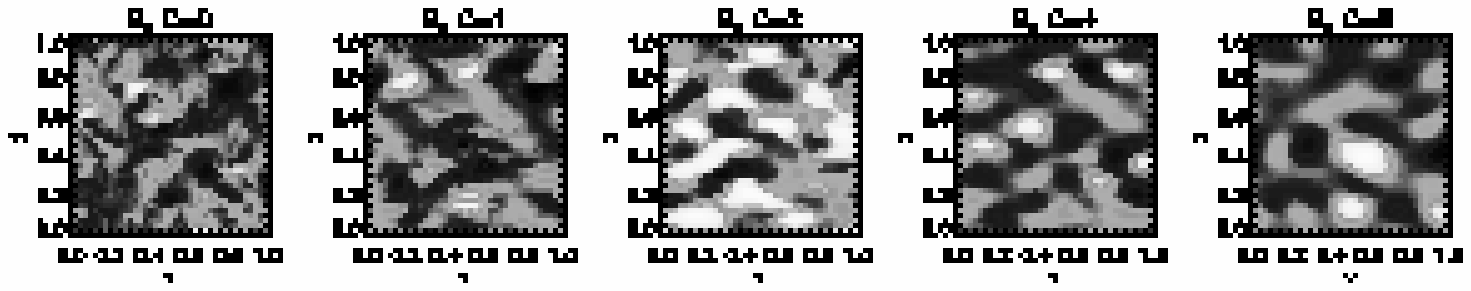}
\caption{Slices of real configuration space $B_x$ fields 
at $t=0.2$ for 3D EMHD turbulence in the presence of a background 
field. From left to right, we show the fields for $C = 0$, 1, 2, 
4 \& 8. In all cases $z = 0.5$.}
\label{background2}
\end{figure}

In order to understand how hyperdiffusivity has affected previous
studies we have introduced a background field into the governing 
equations as discussed above and calculated solutions for $R_B = 100$ 
at a spatial resolution of $256^3$ points in real configuration 
space. We present our results in Figure \ref{background1}. 
>From left to right, we show 2D energy spectra for $R_B = 100$ with 
$C=0$, 1, 2, 4 \& 8. In the isotropic case with no background field, 
i.e. $C=0$, energy is evenly distributed between $x$ and $y$, as 
indicated by circular contours. In the case of $C=1$ we find energy
transfer to larger $k$ has been suppressed in the $x$ direction, 
parallel to the background field. 3D decaying EMHD turbulence has 
become anisotropic in the presence of a uniform background field 
with normal diffusivity. The effect becomes more pronounced for 
$C=2$ and particularly strong for $C=4$ and $C=8$. It is at these
high values of $C$ that the cascade is almost turned 
off in the direction parallel to the field and even strongly suppressed
in the perpendicular directions. In 2D this suppression has been 
attributed to excitation of whistler waves, which act to weaken 
spectral transfer along the direction of propagation 
(\cite[Dastgeer and Zank 2003]{dastgeer03}). In Figure \ref{background2} we show a slice
through the real configuration space field $B_x$ at $z = 0.5$.
For $C=0$, the field is isotropic, but as the value of $C$ is 
increased, structures stretch in the $x$ direction 
corresponding to increasingly inhibited energy transfer in the $x$ direction
but not in $y$. We find the same for $x$ compared to $z$.
At $C=8$, the structures have remained almost the same as at $t=0$
since the cascade in all directions has been so strongly inhibited.

\cite{cho04} performed 3D simulations of EMHD turbulence. They
introduced a uniform background field of strength comparable to 
that of the fluctuating freely decaying field (i.e. $C=1$). 
The authors employed a hyperdiffusivity of 3 and found 
scale-dependent anisotropy. Our simulation at $C=1$ supports their
result. Further, our simulations imply the linear relationship 
between EMHD turbulence and strength of external magnetic field
found by \cite{dastgeer03} in 2D can be extended to 3D.

Note that (\ref{eq:A}) is scale invariant, i.e. it is possible
to apply the equation over the whole of a system, or just to a small 
section, with $R_B$ unchanged.  A very small box then will see the 
large-scale field as a background field, and therefore the smallest 
scales in the system, for example a neutron star, should be 
anisotropic. 

\subsection{Intermediate-scale initial conditions}

In 2D classical turbulence, the exchange of energy and enstrophy 
$\Omega$ is coupled in Fourier space according to
\begin{equation}
\frac{\partial E}{\partial t} = - k^2 \frac{\partial \Omega}
{\partial t},
\end{equation}
hence energy injected at intermediate scales experiences a transfer 
to both higher and lower wavenumbers in order to satisfy this coupling 
and simultaneously conserve energy and enstrophy. This is the inverse 
cascade of energy to lower $k$ (larger scales) (\cite[Kraichnan 1967]{kraichnan67}). In 
MHD turbulence, energy and magnetic helicity are coupled in the same way
and an inverse cascade occurs in order to simultaneously conserve these 
two quadratic ideal invariants. In driven 2D EMHD tubulence, the inverse 
cascade found by \cite{shaikh05} has been proposed to be the result of 
simultaneous conservation of energy and mean squared magnetic potential, 
or anastrophy. Our recent simulations (\cite[WH09]{wareing09}) have shown an
inverse cascade occurs in freely decaying 2D EMHD turbulence. Whilst
it has a considerably different scaling nature to driven 2D EMHD turbulence,
we believe it is also the result of simultaneous conservation of
energy and anastrophy. In 3D then, the existence of an inverse cascade
is immediately under question since anastrophy is no longer conserved.
We inject energy over the wavenumber range $10 \leq k \leq 20$ as 
detailed above and evolve the magnetic field to assess what inverse
cascade there is, if any.

In this case, we have found that simulations with normal diffusivity
are unable to produce an inverse cascade. Since the values of $R_B$
available to us are considerably less than $R_B = 1000$, where decaying
2D EMHD turbulence displays an inverse cascade, this is not a 
surprising result.  For the runs in this section we therefore introduce
a hyperdiffusivity, so that the large scales see a diffusivity equivalent
to $R_B>1000$, whereas the small scales are much more strongly damped, so
that the computations can be done at all. As noted above, using a
hyperdiffusivity will of course disrupt features like the precise shape
of the spectrum at large $k$, including the presence or absence of a
dissipative cutoff. However, since here we are interested in the
behaviour at more modest $k$, in particular the possibility of transferring
energy from intermediate to small wavenumbers, these distortions at large
$k$ probably do not have that much effect on the results.

The precise form of hyperdiffusivity is simply to raise the power of
the diffusion operator,
\begin{equation}
\frac{\partial {\bf B}}{\partial t} =
 - \nabla \times \left[ {\bf J} \times {\bf B} \right] 
- \epsilon \ (\nabla^2)^2 {\bf B},
\label{eq:hyper}
\end{equation}
where $\epsilon$ is the new diffusion coefficient, the equivalent of
$R_B^{-1}$, but only as seen by the largest scales.

\begin{figure}
\includegraphics[angle=0,width=8cm]{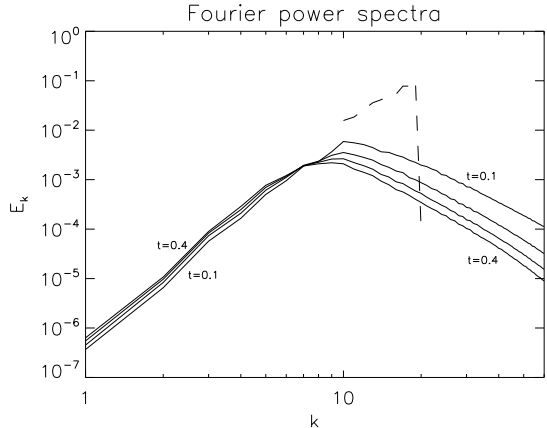}
\caption{Time evolution of the Fourier power spectra for $\epsilon=2.5\cdot10^{-6}$.
Shown are power spectra at $t=0$, 0.01, 0.02, 0.03 and 0.04.  No further inverse
cascase develops after $t=0.04$.}
\label{inverse}
\end{figure}

Figure \ref{inverse} shows spectra
at $t=0$, 0.01, 0.02, 0.03 and 0.04. The spectra show that energy is 
transferred to $k<10$ in a very weak inverse cascade. The spectral peak is
slowly shifting to $k<10$ but not maintaining the same amplitude. 
Some energy has also been transferred to $k>20$. In 3D then, any inverse
cascade is considerably weaker than in 2D, presumably since anastrophy
is no longer conserved.

\section{Conclusions}\label{sec:conc}

We have investigated the nature of decaying 3D EMHD turbulence with 
normal diffusivity and compared it with classical and MHD turbulence 
and studies of 2D and 3D EMHD with hyperdiffusivity. We have found
3D EMHD turbulence experiences an isotropic forward cascade of energy 
to higher wavenumber (smaller spatial scales) asymptotically
approaching $E_k \propto k^{-2}$ with increasing $R_B$ (inversely
proportional to a dissipation coefficient). This is in agreement
with the original theoretical prediction of \cite{goldreich92},
suggesting freely decaying 3D EMHD turbulence is mediated by whistler waves.
Unlike the theoretical prediction, we have found there is no 
dissipative cutoff at the predicted wavenumber $k \sim R_B$ and 
argue this is consistent with non-local coupling in Fourier space, 
the most important result of this paper and consistent with our recent
results for 2D EMHD turbulence. Hyperdiffusivity has previously 
clouded this issue and introduced an artificial cutoff and the unwanted 
bottleneck effect. We have also 
found that fully developed 3D EMHD turbulence appears to be strongly 
structured, retaining a similarity to the initial field at late time, 
very much unlike classical or MHD turbulence.
 
3D EMHD turbulence with normal diffusivity has been found to display 
scale-dependent anisotropy in the presence of a uniform background 
field, in agreement with previous studies employing hyperdiffusivity. 
Our results support previous studies which found the strength of the 
anisotropy is linearly related to the external field strength. Further, 
strong fields effectively halt the cascade in the parallel direction 
and considerably inhibit the cascade in all other directions.

Finally, we have discovered that decaying EMHD turbulence yields a
weak inverse cascade, at least partially transferring magnetic energy 
from intermediate to large lengthscales. The possibility of inverse
cascades may have implications for the magnetic fields of neutron stars,
where the proto-neutron star that emerges from a supernova explosion may
well have a primarily small-scale, disordered field. A Hall-induced inverse
cascade may then be a mechanism whereby it acquires a large-scale, ordered
field.  We note though that the electron number density $n$ is strongly
depth-dependent in neutron stars, which turns out to interact with the
Hall effect in a highly nontrivial way (\cite[Vainshtein et al. 2000, 
Hollerbach and R\"udiger 2004]{vainshtein,hollerbach04}).
It is likely therefore that both forward and inverse cascades will be
rather different in real neutron stars than they are in isotropic,
Cartesian box models such as here.  Future work will consider EMHD in
more realistic, stratified spherical-shell models.

\begin{acknowledgments}
This work was supported by the Science \& Technology 
Facilities Council [grant number PP/E001092/1].
\end{acknowledgments}


\begin{thebibliography}{}

\bibitem[Berger (1997)]{berger97} 
  M.A. Berger 1997
  {\it J. Geophysical Research} {\bf 102}, 2637.

\bibitem[Biskamp, Schwarz and Drake (1996)]{biskamp96} 
  D. Biskamp, E. Schwarz and J.F. Drake 1996 
  {\it Phys. Rev. Lett.} {\bf 76}, 1264.
  
\bibitem[Biskamp et al. (1999)]{biskamp99} 
  D. Biskamp, E. Schwarz, A. Zeiler, A. Celani and J. Drake 1999
	{\it Phys. Plasmas} {\bf 6}, 751.
  
\bibitem[Biskamp and M\"uller (1999)]{biskamp99b} 
  D. Biskamp and W.-C. M\"uller 1999
  {\it Phys. Rev. Lett.} {\bf 83}, 2195-2198.
  
\bibitem[Cho and Lazarian (2004)]{cho04} 
  J. Cho and A. Lazarian 2004
  {\it Astrophys. J.} {\bf 615}, L41.

\bibitem[Dastgeer et al. (2000)]{dastgeer00} 
  S. Dastgeer, A. Das, P. Kaw and P. Diamond 2000
  {\it Phys. Plasmas} {\bf 7}, 571.
  
\bibitem[Dastgeer and Zank (2003)]{dastgeer03} 
  S. Dastgeer and G.P. Zank 2003
  {\it Astrophys. J.} {\bf 599}, 715.

\bibitem[Frigo and Johnson (2005)]{frigo05} 
  M. Frigo and S.G. Johnson 2005 
  {\it Proc. of the IEEE} {\bf 93}, 216.

\bibitem[Galtier (2006)]{galtier06} 
  S. Galtier 2006
  {\it J. Plasma Phys.} {\bf 72}, 721.

\bibitem[Goldreich and Reisenegger (1992)]{goldreich92} 
  P. Goldreich and A. Reisenegger 1992
  {\it Astrophys. J.} {\bf 395}, 250--258.

\bibitem[Hollerbach and R\"udiger (2002)]{hollerbach02} 
  R. Hollerbach and G. R\"{u}diger 2002
  {\it Mon. Not. Roy. Astron. Soc.} {\bf 337}, 216.

\bibitem[Hollerbach and R\"udiger (2004)]{hollerbach04} 
  R. Hollerbach and G. R\"{u}diger 2004
  {\it Mon. Not. Roy. Astron. Soc.} {\bf 347}, 1237.
  
\bibitem[Iroshnikov (1964)]{iroshnikov64}
  P.S. Iroshnikov 1964
  {Sov. Astron.} {\bf 7}, 566

\bibitem[Kolmogorov (1941)]{kolmogorov41} 
  A.N. Kolmogorov 1941 
  {\it Proc. USSR Acad. Sciences} {\bf 30}, 299 (Russian);
  {\it Proc. Roy. Soc. A} {\bf 434}, 9 (1980) (English).

\bibitem[Kraichnan (1965)]{kraichnan65}
  R.H. Kraichnan 1965
  {\it Phys. Fluids} {\bf 8}, 1385

\bibitem[Kraichnan (1967)]{kraichnan67} 
  R.H. Kraichnan 1967
  {\it Phys. Fluids} {\bf 10}, 1417.

\bibitem[Oughton et al. (1998)]{oughton98} 
  S. Oughton, W.H. Matthaeus and S. Ghosh 1998 
  {\it Phys. Plasmas} {\bf 5}, 4235.

\bibitem[Shaikh and Zank (2005)]{shaikh05} 
  D. Shaikh and G.P. Zank 2005
  {\it Phys. Plasmas} {\bf 12}, 122310.

\bibitem[Shebalin et al. (1983)]{shebalin83} 
  J.V. Shebalin, W.H. Matthaeus and D. Montgomery 1983
  {\it J. Plasma Phys.} {\bf 29}, 525.

\bibitem[Vainshtein et al. (2000)]{vainshtein}
  S.I. Vainshtein, S.M. Chitre and A.V. Olinto 2000
  {\it Phys. Rev. E} {\bf 61}, 4422.
  
\bibitem[Wareing and Hollerbach (2009)]{wareing09} 
  C.J. Wareing and R. Hollerbach 2009
  {\it Phys. Plasmas}, in press (WH09)

\end{thebibliography}
\end{document}